\documentclass{article}
\pdfoutput=1
\usepackage{ijcai19}

\usepackage{times}
\usepackage{soul}
\usepackage{url}
\usepackage[utf8]{inputenc}
\usepackage[small]{caption}
\usepackage{graphicx}
\usepackage{amsmath}
\usepackage{booktabs}
\usepackage{algorithm}
\usepackage{algorithmic}
\usepackage{xcolor}
\usepackage{amssymb}
\usepackage{multirow}

\usepackage{adjustbox}
\usepackage{balance}

\DeclareMathOperator*{\argmax}{argmax}

\newcommand{\nosection}[1]{\subsubsection{#1}}

\newcommand{\bnn}{BAYHENN }

\title{BAYHENN: Combining Bayesian Deep Learning and Homomorphic Encryption for Secure DNN Inference}

\author{Peichen Xie$^{1,2}$\thanks{Equal contribution. } \and
Bingzhe Wu$^{1,3}$\footnotemark[1] \And
Guangyu Sun$^{1,2}$
\affiliations
$^1$Center for Energy-efficient Computing and Applications, Peking University\\
$^2$Advanced Institute of Information Technology, Peking University\\
$^3$Ant Financial Services Group\\
\emails
\{xpc, wubingzhe, gsun\}@pku.edu.cn
}

\begin{document}
    \maketitle
    \begin{abstract}
Recently, deep learning as a service~(DLaaS) has emerged as a promising way to facilitate the employment of deep neural networks~(DNNs) for various purposes. 
However, using DLaaS also causes potential privacy leakage from both clients and cloud servers. 
This privacy issue has fueled the research interests on the privacy-preserving inference of DNN models in the cloud service. 
In this paper, we present a practical solution named \bnn for
secure DNN inference. It can protect both the client's privacy and server's privacy at the same time. 
The key strategy of our solution is to combine homomorphic encryption and Bayesian neural networks. Specifically, we use homomorphic encryption to
protect a client's raw data and use Bayesian neural networks
to protect the DNN weights in a cloud server. To verify the
effectiveness of our solution, we conduct experiments on MNIST and a real-life clinical dataset. Our solution achieves consistent latency decreases on both tasks. In particular, our method can outperform the best existing method~(GAZELLE) by about 5$\times$, in terms of end-to-end latency.
\end{abstract}
    \section{Introduction}
\label{sec:intro}
In the past years, deep neural networks~(DNNs) have achieved remarkable
progress in various fields, such as computer vision~\cite{he16,ren17}, 
natural language processing~\cite{vaswani17,devlin18}, and medical image analysis \cite{litjens17,havaei17}. Recently, deep learning as a service~(DLaaS) has emerged as a promising to further enable the widespread use of DNNs in industry/daily-life.
Google\footnote{https://cloud.google.com/inference/}, Amazon\footnote{https://docs.aws.amazon.com/machine-learning/index.html}, and IBM~\cite{bhattacharjee17} have all launched DLaaS platforms in their cloud services. Using DLaaS, a client sends
its private data to the cloud server. Then, the server is responsible
for performing the DNN inference and sends the prediction results back to the client. 
Obviously, if the private client data are not protected, using DLaaS will cause potential privacy issues. A curious server may collect sensitive information contained in the private data (\emph{i.e.} \textbf{client's privacy}). 

To address this privacy issue, researchers have employed the homomorphic encryption to perform various DNN operators on encrypted client data~\cite{phong18,ma19}. As a result, the cloud server only serves as a computation platform but cannot access the raw data from clients. 
However, there exist two major obstacles in applying these approaches. First, some common non-linear activation functions, such as ReLU and Sigmoid, are not cryptographically computable. Second, the inference processing efficiency is seriously degraded by thousands of times. 

To tackle these problems, a recent work~\cite{zhang18} proposes using an interactive paradigm. A DNN inference is partitioned into linear and non-linear computation parts. Then, only the linear computations are performed on the cloud server with encrypted data. The non-linear computations are performed by the client with raw data. However, in such an interactive paradigm, the intermediate features extracted by the linear computations are directly exposed (sent back) to the client. Thus, a curious client can leverage these features to reconstruct the weights of the DNN model held by the cloud~\cite{tramer16,zhang18}. This issue is called the leakage of \textbf{server's privacy}.

In fact, a practical solution for secure DNN inference should protect both client's privacy and server's privacy. In addition, it should support DNNs with all types of non-linear activation functions. Unfortunately, there still lacks an effective approach in literature. The limitations of recent works are discussed in Section \ref{sec:lim}. Thus, we propose a new method for secure DNN inference, called BAYHENN. Our key strategy is to combine Bayesian deep learning and homomorphic encryption. To the best of our knowledge, it is the first approach that can protect both client's privacy and server's privacy and support all types of non-linear activation functions at the same time. 

BAYHENN follows an interactive paradigm so that all types of activation functions are supported. On the one hand, it uses the homomorphic encryption to provide protection for the client's privacy. On the other hand, it adds proper randomness into DNN weights to prevent the leakage of the server's privacy. This idea is motivated by the learning with error~(LWE) problem, in which solving the noisy linear equations system has been proven
to be NP-hard \cite{regev09}. However, \emph{``adding proper randomness''} is a new challenge. Directly injecting well-designed Gaussian Noise into weights will cause a drastic performance degradation~\cite{chaudhuri08,zhang17}. Therefore, we propose to use a more intrinsic method, the Bayesian neural network, to model the weight uncertainty of DNNs. Armed with the weight uncertainty, we can protect the weight information by sending obscured features to the client.

In summary, the contributions of our paper are as follows:
\begin{itemize}
    \item We provide a novel insight that using the Bayesian neural network can prevent leakage of a server's privacy under an interactive paradigm.
    \item Based on this insight, we build a practical solution for secure DNN inference. Our solution is more (about 5$\times$) efficient than the best existing method, and is capable of supporting all types of activation functions.
    \item We validate the proposed solution on a real-life clinical dataset, which is less explored in previous studies.
\end{itemize}

    \section{Preliminary}
\label{sec:pre}
In this part, we first introduce some basic notations in
DNN inference. Then, the potential privacy leakages in DLaaS and the goal of this paper are
presented. At last, we summarize the limitations of previous works on secure DNN inference.

\subsection {DNN Inference}
We start from the process of a DNN inference. For simplicity, the following discussion
is based on a fully connected neural network. 
It can be easily switched to the context of a convolutional neural network, as shown in our experiments. 

We take the image classification as an example in this part. It can be extended to other types of tasks as well~(\emph{e.g.} regression). The image classification aims to assign a label $t$ to the input image $\mathbf{x}$. To this end, 
a classification model $\mathbb{M}: {t}=f\left(\mathbf{x};\boldsymbol{\theta}\right)$ is trained on some pre-collected data. In a context of deep learning,
the $f$ is a highly non-linear function, which is formed by alternately stacking several linear~(\emph{e.g.} fully connected layer) and non-linear~(\emph{e.g.} activation function) operations\footnote{ Most previous DNNs satisfy this topology. Note that one can easily integrate two adjacent linear/nonlinear layers into one ``logic" layer. For example, we can combine one batch normalization and one convolution layer together.}. 

Formally, for a neural network, we denote $\mathbf{W}^{(i)}$ and $\mathbf{b}^{(i)}$ as weights and biases corresponding to the $i$-th linear layer.
To calculate outputs~(\emph{i.e.} activations) of the $i$-th non-linear layer~(denoted as $\mathbf{a}^{(i)}$), a linear transformation is firstly performed on the inputs from previous layers following:
\begin{equation}
\label{equ:linear}
\mathbf{z}^{(i)} = \mathbf{W}^{(i)}\mathbf{a}^{(i-1)} + \mathbf{b}^{(i)}
\end{equation}
Then, the nonlinear activation function is used to obtain output activation results: 
\begin{equation}
\label{equ:nonlinear}
\mathbf{a}^{(i)} = \varphi^{(i)}(\mathbf{z}^{(i)})
\end{equation}
where $\varphi^{(i)}$ denotes the
activation function~(\emph{e.g.} ReLU). 

In summary, the classification model 
$\mathbb{M}$ can be
reformulated into:
\begin{equation}
    \mathbb{M}: t = f (\mathbf{x};\{(\mathbf{W}^{(i)},\mathbf{b}^{(i)})\}^{n}_{i=1})
\label{equ:dnn_inferecne}
\end{equation}
\subsection{Privacy Leakage in DLaaS}
\label{sec:leak}
In a common setting, there are two parties involved in a process of DLaaS. Specifically, at the beginning, a \textit{client} sends its
private data $\mathbf{x}$ into a cloud \textit{server}, which holds the DNN model $\mathbb{M}$ for inference. Then, the server can perform the DNN inference and send the prediction results back to the client.
Considering that both the client and the server are semi-honest, privacy leakages may occur on both sides:
\begin{itemize}
    \item \textbf{Client's Privacy Leakage:} From the perspective of a client, its input data $\mathbf{x}$ is directly exposed to the cloud server in a
    non-private setting. Thus, the server may collect the sensitive information contained in $\mathbf{x}$.
    \item \textbf{Server's Privacy Leakage:} A cloud server also holds the private data, \emph{i.e.} the weight parameters and the DNN structure of $\mathbb{M}$. A curious client may attempt to obtain these valuable model information.
\end{itemize}
In this paper, we focus on a secure DNN inference. Our goal is to
prevent the privacy leakage of the input data~(client-side) and
the model parameters~(server-side). The protection of a model structure is out of the scope of this paper, yet part of the DNN model structure (such as the filter and stride size in the convolution layers and the type of each layer) is also protected in our solution\footnote{Juverka \emph{et al.} \shortcite{juvekar18} have explained how homomorphic encryption can hide this information.}. 

\subsection{Limitation of Previous Work}
\label{sec:lim}
In this subsection, we discuss the limitations of previous works, which are most related to ours (listed in Table~\ref{tab:lim}). We do not consider \cite{bourse18,sanyal18}, since they only work on binary neural networks and do not outperform the state-of-the-art.

\begin{table}
\centering
\begin{tabular}{@{}cccc@{}}
\toprule
\textbf{}  & \begin{tabular}[c]{@{}c@{}}Client's\\  Privacy \end{tabular} & \begin{tabular}[c]{@{}c@{}}Server's\\ Privacy\end{tabular} & \begin{tabular}[c]{@{}c@{}}Support all\\ activation func.\end{tabular} \\ \midrule
CryptoNets & Yes & Yes & --- \\
GELU-Net & Yes & --- & Yes \\
GAZELLE & Yes & Yes & --- \\
Ours & Yes & Yes & Yes \\ \bottomrule
\end{tabular}
\caption{Feature comparison among different solutions.}
\label{tab:lim}
\end{table}

CryptoNets \cite{gilad-bachrach16} is the first system for homomorphic encryption based neural network inference.
However, its end-to-end latency for a single input is extremely high, especially on a DNN. More importantly, CryptoNets cannot support most of the common activation functions, such as Sigmoid and ReLU, or the common pooling functions, such as Max-Pooling. These two issues limit the use of CryptoNets in real-life scenarios.

GELU-Net \cite{zhang18} proposes using an interactive paradigm to address the issues of CryptoNets. 
However, the interactive paradigm may result in server's private weight leakage. Thus, GELU-Net needs to limit the number of queries requested by one specific client.

GAZELLE \cite{juvekar18} is state-of-the-art work combining garbled circuits and homomorphic encryption. Garbled circuits can be used to protect the client's weights \cite{liu17,mohassel17,rouhani18}. However, such a method introduces non-negligible computation/communication costs, and does not support those activation functions which are garbled-circuit unfriendly (\emph{e.g.} Sigmoid).

Compared with these works, our proposed
solution can overcome all the above limitations. Moreover, unlike these works, we validate our solution on a real-life clinical dataset, which further shows the practicability of our solution in real-life applications. 

    \section{Our Approach}
In this section, we first describe the high-level protocol of BAYHENN and then detail its implementation for secure linear/non-linear computations.

\subsection{Protocol at a High Level}
Here, we elaborate our protocol for secure DNN inference at a high level.
The core idea for designing our protocol is based on the combination of homomorphic encryption
and deep Bayesian inference. 

At a high level, our protocol follows an interactive paradigm and comprises of two sub-protocols, namely, the SLC~(\textbf{S}ecure \textbf{L}inear \textbf{C}omputation) and the SNC~(\textbf{S}ecure \textbf{N}on-linear \textbf{C}omputation) protocols. Consider a basic setting where the centric server holds the model $\mathbb{M}$ and the client
holds the private data $\mathbf{x}$. The inference task $t=f(\mathbf{x};\boldsymbol{\theta})$ is 
decomposed into linear and non-linear computations based on Equation~(\ref{equ:linear}) and (\ref{equ:nonlinear}). Each of them is performed by different parties~(the server and the client) using the SLC and SNC protocols, respectively. 
To be specific, the client firstly encrypts their private data
$\mathbf{x}$ and sends the encrypted data to the server as model input. Then, the
server can perform the linear computations on the encrypted input following the SLC protocol. Due to most commonly used
non-linear activation functions are not cryptographically computable, the output is sent back to the client for the further non-linear computation following the SNC protocol. Then, the
client can re-encrypt the computed results and send them to the server for computation of the next layer. Once all computation tasks are finished, the client can receive the final prediction $t$. The whole procedure of our protocol is depicted
in Algorithm~\ref{algo:DNN}. Two core sub-protocols, SLC and SNC, are detailed in the following part.

\begin{algorithm}[t]
\caption{Secure DNN inference}
\label{algo:DNN}
\textbf{Input}: $(\mathbf{x};\  S, P(\boldsymbol{\theta}|\mathcal{D}), \{\varphi^{(i)}\}^{n}_{i=1})$\\
\textbf{Output}: $(t;\  \varnothing)$
\begin{algorithmic}[1]
\STATE Initiation
\STATE Server:
\FOR{$k \gets 1$ \TO $S$}
\STATE samples $\boldsymbol{\theta}_k=\{\mathbf{W}^{(i)}_k, \mathbf{b}^{(i)}_k\}_{i=1}^n$ from $P(\theta|\mathcal{D})$
\ENDFOR
\STATE Server: $\Theta \gets \{\boldsymbol{\theta}_k\}_{k=1}^S$ 
\STATE Client: $\mathbf{a}^{(0)} \gets \mathbf{x}$
\FOR{$i\gets 1$ \TO $n$}
\STATE $\tilde{\mathbf{z}}^{(i)} \gets \mathrm{SLC}(\mathbf{a}^{(i-1)};\ \Theta)$
\STATE $\mathbf{a}^{(i)} \gets \mathrm{SNC}(\tilde{\mathbf{z}}^{(i)};\ \varphi^{(i)})$
\ENDFOR
\STATE Client: $\mathbf{p} \gets \frac{1}{S} \sum_{i=1}^{S}\mathbf{a}^{(n)}_k$
\STATE Client: $t \gets \argmax_k p_k$
\end{algorithmic}
\end{algorithm}

\subsection{SLC Protocol}
The aim of this protocol is to enable the linear computations~(\emph{e.g.} convolution) while providing the protection for
both the client's data and the server's parameters.
In the following parts, we first introduce some basic functions in a vectorizable homomorphic encryption scheme. Then, we demonstrate how SLC protocol 
can successfully achieve the server-level~(server's parameters) and client-level~(client's data)
privacy protection. 

\nosection{Vectorizable Homomorphic Encryption}
To ensure the correctness of the linear operations in ciphertext, we adopt homomorphic encryption, which provides three basic functions: the encryption function $\mathbb{E}$, the decryption function $\mathbb{D}$ and the evaluation function $\mathrm{Eval}$. 
To be specific, $\mathbb{E}$ is responsible for encrypting the plaintext $x$ to the ciphertext $\tilde{x}$ under a public key. In contrast, $\mathbb{D}$ is to decrypt the ciphertext into the plaintext under a secret key. The homomorphic operations are instantiated by the evaluation function. For two elements $x_1, x_2 \in \mathcal{R}$~($\mathcal{R}$ denotes the plaintext space as a ring), we can establish the following equations:
\begin{gather}
    \mathbb{D}(\mathrm{Eval}(\mathbb{E}(x_1), \mathbb{E}(x_2), +)) = x_1 + x_2 \label{equ:add}\\
    \mathbb{D}(\mathrm{Eval}(\mathbb{E}(x_1), x_2, \times)) = x_1 \times x_2 \label{equ:mul}
\end{gather}
where $+$ and $\times$ denote the normal addition and multiplication in the plaintext space $\mathcal{R}$, respectively. For simplicity, we will use the notation $\tilde{x}_1 \oplus \tilde{x}_2$ and $\tilde{x}_1 \otimes x_2$ in the following part to represent the instantiation of the ciphertext-ciphertext addition $\mathrm{Eval}(\mathbb{E}(x_1), \mathbb{E}(x_2), +)$ and the ciphertext-plaintext multiplication $\mathrm{Eval}(\mathbb{E}(x_1), x_2, \times)$ respectively. Note that the ciphertext-ciphertext multiplication is not needed in our approach.

It is impractical and inefficient to directly apply a conventional homomorphic encryption scheme to the DNN inference. Firstly, the above functions are defined over the ring $\mathcal{R}$ whereas the parameters in a DNN model are typically floating points. 
Then, operations~(\emph{e.g.} convolution) in a DNN model are always performed on the high-dimensional tensors, which leads to the inefficiency of the straightforward solution~(\emph{i.e.} to compute the tensor multiplication/addition element by element). 
To tackle these two issues, we adopt a \emph{vectorizable} homomorphic encryption scheme\footnote{The vectorization is also know as ``batching'' or ``SIMD'' in the literature.}, which involves the $\mathrm{Encode}$ function. This function is capable of packing a group of floating points to a single element in $\mathcal{R}$~(\emph{i.e.} one plaintext). On the contrary, the $\mathrm{Decode}$ function is to transform the plaintext into a number of floating points.
Specifically, we implement the encoding and decoding functions following prior works~\cite{gilad-bachrach16,juvekar18}.

\nosection{Client-side Privacy Protection}
To protect the privacy of the client's data, we make use of homomorphic encryption. Homomorphic encryption enables the client to encrypt the private data and the server to conduct the linear computation on the encrypted data. Here, we use notations in Section~\ref{sec:pre} to explain how a vectorizable homomorphic encryption scheme works in DNN inference.
To compute the output of the $i$-th linear layer ($\mathbf{z}^{(i)}$), the client 
first encrypt the $i$-th input vector $\mathbf{a}^{(i-1)}$ into $\tilde{\mathbf{a}}^{(i-1)}$ via the encoding and
encryption functions. For the linear computation, we can always decompose it into a number of multiply-accumulate operations of vectors. Thus we can use the Equation~\ref{equ:add} and \ref{equ:mul} and the above vectorization technique to achieve the computation of the linear layer.  
For simplicity, we formulate the process of the linear computation as\footnote{Here, we take FC layer as an example, in practice, it is natural to extend such a process to the convolution layer.}:
\begin{equation}
\label{equ:mac}
    \tilde{\mathbf{z}}^{(i)} = \mathrm{Encode}(\mathbf{W}^{(i)})\otimes \tilde{\mathbf{a}}^{(i-1)}\oplus \mathbb{E}(\mathrm{Encode}(\mathbf{b}^{(i)}))
\end{equation}
With a little abuse of notations, we use $\otimes$
to represent the multiplication between an unencrypted matrix and an ecrypted vector. 
Armed with the above equations, the server can achieve linear computation without knowing any information about the client's data.

\nosection{Server-side Privacy Protection} As discussed in Section~\ref{sec:leak}, the interactive paradigm comes with the risk of the weight leakage.
To tackle this issue, motivated by the learning with error (LWE)
problem~\cite{regev09}, we propose using Bayesian neural networks to build the model parameters with
moderate uncertainty.

The BNN is to model the weight as the probability
distributions instead of the fixed floating points in the
traditional DNN. Here, we borrow some notations from the DNN inference. The target of the BNN is to predict 
the label distribution $P(t|\mathbf{x})$. Given the training dataset
$\mathcal{D}$, we can estimate the posterior distribution of the weights $P(\boldsymbol{\theta}|\mathcal{D})$ via Bayes
by Backprop~\cite{blundell15}. Then, the posterior distribution can be used for label prediction. Specifically,
given a test data $\mathbf{x}$ (client's data in our case),
the label distribution can be obtained following:
\begin{equation}
    P(t|\mathbf{x}) = \mathbb{E}_{\boldsymbol{\theta}\sim P(\boldsymbol{\theta|\mathcal{D}})}(P(t|\mathbf{x}, \boldsymbol{\theta}))
\label{equ:BNN_inferce}
\end{equation}

To compute the expectation defined in Equation~\ref{equ:BNN_inferce}, we can sample a number of models based on the posterior distribution. These models share
the same model architecture while having different parameters
sampled from~$P(\boldsymbol{\theta}|\mathcal{D})$. Then,
we can perform the inferences of these models and average the
outputs for the final prediction.
In summary, the whole process can be represented as:
\begin{equation}
    P(t|\mathbf{x})~\approx~\dfrac{1}{S}\sum_{k=1}^{S}
    P(t|\mathbf{x},\boldsymbol{\theta}_{k})
    \label{equ:sample}
\end{equation}
where $S$ denotes the number of sampling. $\boldsymbol{\theta}_k$ is the $k$-th sampling result and
inherits the formulation from Equation \ref{equ:dnn_inferecne}
in which $\boldsymbol{\theta}_k = \{\mathbf{W}^{(i)}_k, \mathbf{b}^{(i)}_k\}_{i=1}^n$. The computation of Equation \ref{equ:sample} can be treated as the computation of $\{P(t|\mathbf{x},\boldsymbol{\theta}_{k})\}_{k=1}^S$, which is naturally parallelizable. Since the modern server easily allows process-level parallelism, the computation can be seen as $S$ parallel processes of DNN inference and do not significantly increase the computation time.

During the inference of a BNN, due to the features exposed to the client are inaccurate compared with the normal DNN inference, it is hard for the client to solve the exact weights of the server. 
The hardness comes from the challenge of the LWE problem, where solving the noisy linear system is as difficult as solving lattice problems\footnote{The lattice problems are used as a base to build homomorphic encryption schemes \cite{acar18}.}.
Note that the weights of the BNN are
the parameters of massive
probability distributions. In our case, each $\mathbf{W}^{(i)}$
follows the multivariate Gaussian distribution, and the weights of the layer are the mean and variance vectors of the Gaussian
distribution.

Put all things together, we can build the SLC protocol,
which can prevent the leakage of weights and input data
simultaneously. The SLC protocol is depicted in Algorithm~\ref{algo:linear}.

\begin{algorithm}[t]
\caption{SLC}
\label{algo:linear}
\textbf{Input}: $(\mathbf{a}^{(i)};\  \mathbf{W}^{(i)}, \mathbf{b}^{(i)})$\\
\textbf{Output}: $(\mathbf{z}^{(i)};\  \varnothing)$
\begin{algorithmic}[1]
\STATE Client: $\tilde{\mathbf{a}}^{(i)} \gets \mathbb{E}(\mathrm{Encode}(\mathbf{a}^{(i)}))$
\STATE Client: sends $\tilde{\mathbf{a}}^{(i)}$ to the server
\STATE Server: starts $S$ threads
\FOR{$k\gets 1$ \TO $S$}
\STATE $\tilde{\mathbf{z}}^{(i)}_k \gets \mathrm{Encode}(\mathbf{W}^{(i)}_k)\otimes \tilde{\mathbf{a}}^{(i-1)}_k\oplus \mathbb{E}(\mathrm{Encode}(\mathbf{b}^{(i)}_k))$
\ENDFOR
\STATE Server: $\tilde{\mathbf{z}}^{(i)} \gets \{\mathbf{z}^{(i)}_k\}_{k=1}^S$
\STATE Server: sends $\tilde{\mathbf{z}}^{(i)}$ to the client
\end{algorithmic}
\end{algorithm}

\begin{algorithm}[t]
\caption{SNC}
\label{algo:nonlinear}
\textbf{Input}: $(\tilde{\mathbf{z}}^{(i)};\  \varphi^{(i)})$\\
\textbf{Output}: $(\mathbf{a}^{(i)};\  \varnothing)$
\begin{algorithmic}[1]
\STATE Server: sends $\varphi^{(i)}$ to server
\STATE Client:
\FOR{$k \gets 1$ \TO $S$}
\STATE $\mathbf{z}^{(i)}_k \gets \mathrm{Decode}(\mathbb{D}(\tilde{\mathbf{z}}^{(i)}_k))$
\STATE $\mathbf{a}^{(i)}_k \gets \varphi^{(i)}(\mathbf{z}^{(i)}_k)$
\ENDFOR
\STATE Client: $\mathbf{a}^{(i)} \gets \{\mathbf{a}^{(i)}_k\}_{k=1}^{S}$ 
\end{algorithmic}
\end{algorithm}

\subsection{SNC Protocol}
This protocol is to enable secure non-linear computations of DNNs. In our protocol, the non-linear computations are performed
on the unencrypted data and are executed on the client individually. Specifically, to compute the output of the $i$-th non-linear layer ($\mathbf{a}^{(i)}$), the client first decrypts the input  $\tilde{\mathbf{z}}^{(i)}$ into $\mathbf{z}^{(i)}$ via the decryption function and the decoding function. In the context of Bayesian inference, we will get $S$ vectors of floating points. Then, following Equation \ref{equ:nonlinear}, the client can apply the activation function $\varphi^{(i)}$ to each of these vectors $\mathbf{z}^{(i)}_k$ simultaneously. In summary, the
SNC protocol is described in Algorithm~\ref{algo:nonlinear}.
In the execution process of the SNC protocol, the client
does not need to send any private information to the server, and the server does not expose the weight
parameters explicitly/implicitly to the client. 

    \section{Experiment}
\subsection{Implementation Details}
Considering efficiency and comparability, we instantiate our solution \bnn using the SEAL library\footnote{https://github.com/microsoft/SEAL}, the homomorphic encryption library widely used by previous works. 
For a fair comparison, we adopt SEAL's BFV scheme, which is also used by the state-of-the-art solution GAZELLE. In this scheme, we set the security level to 128-bit, the degree of the polynomial modulus to 2048, the plaintext modulus to 20-bit, and the noise standard deviation to 4.

We deploy \bnn and previous works on a centric server with an Intel Xeon E5-2650 v3 2.30 GHz CPU and a client PC with an Intel Core i3-7100 3.90 GHz CPU. The bandwidth between the server and the client is 100 Mbps. 
To evaluate efficiency, we use the end-to-end latency (including protocol setup and communication) as a metric to measure the performance of different solutions.

\subsection{Performance Evaluation} \nosection{Model Setup} For each DNN architecture used in
our experiments, we need to train three individual versions
to meet the requirements of different secure DNN
inference methods. We list these three versions as follows:
\begin{itemize}
    \item {\tt Normal}: Used by GAZELLE and GELU-Net.
    \item {\tt Square}: Replacing all the activation functions with the square function and replacing all the pooling functions with the sum pooling. Used by CryptoNets.
    \item {\tt Bayes}: Bayesian version of the base DNN, which is trained by Bayes By Backprop. Used by BAYHENN.
\end{itemize}

\nosection{Digit Classification}
The first task we consider is the hand-written digit classification. We build our model based on the
MNIST dataset, which consists of 60,000 training and 10,000 validation images of 28 by 28 pixels. We adopt the LeNet5 \cite{lecun98}
as the basic network architecture with the input size of 28$\times$28.
We select the optimizer following the prior work~\cite{shridhar18}. 
Specifically, we make use of Adam for {\tt Bayes} and {\tt Square}. The learning rate is set to 0.001. For {\tt Normal}, we select
the Momentum SGD as the optimizer. The learning rate and momentum are set to 0.01 and 0.9, respectively. 
For a fair comparison, all these versions are trained without data augmentation or Dropout. The sampling number of {\tt Bayes} is
set to 4. 

The overall results are shown in Table~\ref{tab:mnist}. From the results, our solution achieves the best latency in
contrast to the other three solutions. 
For the classification 
performance, there is a slight accuracy decrease from
{\tt Bayes}~(used by our solution) to {\tt normal}. We infer
this decrease may be caused by that the first-order
optimizer can not efficiently optimize the Bayesian neural
network. On the other hand, both {\tt Bayes} and {\tt Normal}
can outperform {\tt Square} by a large margin. This can be caused by the gradient vanish issue in DNN training, which is brought by the square activation function.
In terms of efficiency, our solution outperforms 
the state-of-the-art work (GAZELLE) by 4.93~s in latency (a speedup of 4.63$\times$) with an accuracy drop of 0.12\%.

\begin{table}
\begin{adjustbox}{width=\columnwidth,center}
\begin{tabular}{@{}ccccc@{}}
\toprule
Framework  & Version      & Scheme   & Accuracy~(\%) & Latency (s) \\ \midrule
CryptoNets & {\tt Square} & YASHE'   & $96.09$       & $3593.22$   \\
GELU-Net   & {\tt Normal} & Paillier & $99.05$       & $107.49$    \\
GAZELLE    & {\tt Normal} & BFV      & $99.05$       & $6.29$      \\
Ours       & {\tt Bayes}  & BFV      & $98.93$       & $1.36$      \\ \bottomrule
\end{tabular}
\end{adjustbox}
\caption{Performance comparison on digit classification.}
\label{tab:mnist}
\end{table}

\begin{table}
\begin{adjustbox}{width=\columnwidth,center}
\begin{tabular}{@{}ccccc@{}}
\toprule
Framework  & Version      & Scheme   & Accuracy~(\%) & Latency (s) \\ \midrule
CryptoNets & {\tt Square} & YASHE'   & $81.66$       & ---         \\
GELU-Net   & {\tt Normal} & Paillier & $83.58$       & $4755.59$   \\
GAZELLE    & {\tt Normal} & BFV      & $83.58$       & $21.64$     \\
Ours       & {\tt Bayes}  & BFV      & $83.36$       & $4.17$      \\ \bottomrule
\end{tabular}
\end{adjustbox}
\caption{Performance comparison on breast cancer classification.}
\label{tab:idc}
\end{table}

\nosection{Classification on Breast Cancer} To further demonstrate the effectiveness of our approach, we apply it to a publicly available dataset for invasive ductal carcinoma (IDC) classification\footnote{ http://www.andrewjanowczyk.com/use-case-6- invasive-ductal-carcinoma-idc-segmentation/}, which contains 277,524 patches of 50$\times$50 pixels (198,738 IDC-negative and 78,786 IDC-positive). 
We chose a modified version of AlexNet~\cite{shridhar18} as the base network architecture. For preprocessing, the input image is resized to
32$\times$32. We adopt the similar training strategy of digit classification, \emph{i.e.} training without data augmentation or Dropout.

The overall results\footnote{The computation of AlexNet using CryptoNets cannot be completed within an acceptable time.} are shown in~Table~\ref{tab:idc}. From
the results, \bnn has achieved consistent improvements on the task of IDC classification. In terms of end-to-end latency, \bnn
significantly outperforms GAZELLE as well as other two previous works.   
In particular, \bnn can outperforms GAZELLE by 17.47~s, which shows a 5.19$\times$
speedup. In contrast to GELU-Net, our method
can speed up for 1140$\times$. This drastic improvement
can be attributed to the use of a more advanced homomorphic encryption scheme.

All the above results indicate that our solution can significantly speed up the secure inference of the DNN compared with previous works. Moreover, as discussed previously, \bnn can provide the protection
of the server's weights without limiting the number of requests
by one client.

\subsection{Discussion}
In this section, we provide some discussions from
the following aspects.

\nosection{Regularization of BNN} In addition to
validating on the testing dataset, we also perform
the evaluation on the training dataset to explore
the regularization effect of the Bayesian neural network.
The results are shown in Table~\ref{tab:gap}.
Here we use the gap between training and testing accuracy
to quantitatively measure the degree of overfitting.
From the results, we observe that the {\tt Bayes} version has
achieved the smallest gap.
For the task of digit classification, {\tt Bayes} decreases
the gap from 0.74 to 0.04 compared with {\tt normal}, while
there is a gap decrease of 2.24 in the task of IDC classification.
We infer the regularization effect comes from the penalty term~(the KL
distance between the prior and variational distribution) in the objective function~(more details can be found in the prior work~\cite{blundell15}).  
These results indicate that the Bayesian neural network leads to a stronger regularization effect than the
traditional methods. Moreover, this fact shows our solution can
be used in the scenario where the training data is scarce.

\begin{table}
\centering
\begin{tabular}{@{}ccccccc@{}}
\toprule
\multirow{2}{*}{Version} & \multicolumn{3}{c}{MNIST}  & \multicolumn{3}{c}{IDC} \\ \cmidrule(l){2-7} 
                         & Train   & Test    & Gap    & Train    & Test  & Gap  \\ \midrule
{\tt Normal}             & $99.79$ & $99.05$ & $0.74$ & $86.63$  &  $83.58$      & $3.05$      \\
{\tt Square}             & $97.87$ & $96.09$ & $1.78$ & $82.67$ & $81.66$      & $1.01$     \\
{\tt Bayes}              & $98.97$ & $98.93$ & $0.04$ & $84.17$ & $83.36$       & $0.81$     \\ \bottomrule
\end{tabular}
\caption{The gap between training and testing accuracy~(\%).}
\label{tab:gap}
\end{table}

\nosection{Communication Cost}
As shown in Table \ref{tab:commu}, the communication cost can be a non-negligible component of the end-to-end latency. Among all the tested frameworks, GAZELLE is most affected by the communication cost (more than 90\% of the latency comes from data transmission). We point out that this is mainly due to the garbled circuits of ReLU and MaxPool, whose size is proportional to the size of intermediate features of the DNN. On the contrary, without the use of garbled circuits, \bnn makes a remarkable decrease in data transmission, which leads to 4.63$\times$ and 5.19$\times$ speedup for LeNet5 and AlexNet respectively.
In addition, the communication cost of GELU-Net is even lower. Indeed, we do need to transfer multiple intermediate features in ciphertext due to the multiple sampling, but this is a reasonable trade-off between communication costs and privacy protection. 

\begin{table}
\centering
\begin{tabular}{@{}ccc@{}}
\toprule
Framework  & LeNet5 on MNIST & AlexNet on IDC \\ \midrule
CryptoNets & $595.50$        & ---            \\
GELU-Net   & $0.83$          & $2.48$         \\
GAZELLE    & $77.95$         & $252.60$       \\
Ours       & $0.81$          & $6.32$         \\ \bottomrule
\end{tabular}
\caption{The communication costs~(MB).}
\label{tab:commu}
\end{table}

\nosection{Other Homomorphic Encryption Schemes}
For a fair comparison, we make use of the BFV scheme in our solution. In fact, we have also conducted a series of experiments to test the performance of other homomorphic encryption schemes including YASHE' and CKKS (vectorizable), and Paillier (not vectorizable). 
Among these schemes and BFV, there is no single scheme that can outperform others for any DNN. Specifically, our solution with the CKKS scheme has a latency of 1.61~s and 10.07~s (with 1.38~MB and 7.75~MB communication costs) for LeNet5 and AlexNet respectively (in addition, YASHE' and Paillier are much slower). This result indicates that the BFV scheme is more efficient for the above benchmarks. 
However, we also noted that CKKS will outperform BFV when the shape of intermediate features enlarges (\emph{e.g.} to 64$\times$64). Therefore, the best scheme should be chosen according to the shape and size of the features.

\nosection{Extension on RNN}
In the above experiments, we have evaluated the effectiveness
of our solution on CNN based models. However, it is natural
to enable the secure inference of recurrent neural
networks~(RNNs). There have been some prior works that
focus on training a Bayesian RNN. A recent work~\cite{fortunato17} has
built a Bayesian RNN with LSTM cells. This model is used for language modeling task and has achieved comparable performance
compared with a normal RNN. All basic operations in the Bayesian
RNN model are included in the scope of our solution. Thus we can
easily integrate the model into the protocol in Algorithm~\ref{algo:DNN}.
In the context of Bayesian RNN, GAZELLE cannot provide the support for the RNN model like our solution. This
is because the computation of the Tanh/Sigmoid functions~(exist in an LSTM cell) is particularly expensive under the garbled circuit protocol.
    \section{Conclusion}
In this paper, we introduce BAYHENN, a practical solution for secure DNN inference.
The key innovation lies in the strategy to combine Bayesian deep learning and homomorphic encryption.
Armed with this strategy, our solution is capable of achieving secure inference
of DNN models with arbitrary activation functions, while our solution enjoys 5$\times$ speedup in contrast to the best
existing work. Applying this method in the DLaaS scenario is promising
and implies a wide range of real-life applications.
This research also points out a new direction, to apply Bayesian deep learning on the tasks about privacy protection.
For example, previous works that focus on training a DNN
model under differential privacy,  can benefit from the
insight in our paper. 

Despite the superiority of our method, there is much room for further improvement. From the view of optimization, how to design a better algorithm to optimize the Bayesian neural network is crucial
to improving the model accuracy. Another direction is to use some
hardware devices~(\emph{e.g.} FPGAs) to accelerate the computation of secure
DNN inference. 

    \section*{Acknowledgements}
    This work was supported by National Natural Science Foundation of China (No.61572045).
    \newpage
    \balance
	\bibliographystyle{named}
    \bibliography{ijcai19}
\end{document}